\documentclass[twocolumn,letterpaper,10pt]{article}
\usepackage{kuz}
\usepackage{graphicx}				
\usepackage[utf8]{inputenc}  				
\usepackage[round,authoryear]{natbib}
\usepackage[labelfont=bf]{caption}			
\usepackage[colorlinks=false]{hyperref} 	

\usepackage{background}
\backgroundsetup{
  position=current page.north,
  angle=0,
  nodeanchor=north,
  vshift=14mm,
  opacity=0.5,
  scale=1,
  contents= {\shortstack{PUBLISHED AS:  Hoffmann, M. (2022), Body schema or the body as its own best model,\\ in G. \v{S}ejnov\'{a} and M. Vavre\v{c}ka and J. Hvoreck\'{y}, ed., `Kognice a um\v{e}l\'{y} \v{z}ivot XX (Cognition and Artificial Life XX)',\\ CTU in Prague, pp. 45-51.}}
}

\makeatletter
\def\convertto#1#2{\strip@pt\dimexpr #2*65536/\number\dimexpr 1#1}
\makeatother

\begin{document}

\date{}
\title{Body schema or the body as its own best model}

\author{
\vspace{1ex}
\textbf{Matej Hoffmann} \\ 
Department of Cybernetics, Faculty of Electrical Engineering, Czech Technical University in Prague\\
Email: matej.hoffmann@fel.cvut.cz \\
}

\maketitle 

\thispagestyle{empty}

\noindent
{ \begin{center} \bf\normalsize Abstract\end{center}}
{
\noindent
Rodney Brooks (1991) put forth the idea that during an agent's interaction with its environment, representations of the world often stand in the way. Instead, using the world as its own best model, i.e. interacting with it directly without making models, often leads to better and more natural behavior. The same perspective can be applied to representations of the agent's body. I analyze different examples from biology---octopus and humans in particular---and compare them with robots and their body models. At one end of the spectrum, the octopus, a highly intelligent animal, largely relies on the mechanical properties of its arms and peripheral nervous system. No central representations or maps of its body were found in its central nervous system. Primate brains do contain areas dedicated to processing body-related information and different body maps were found. Yet, these representations are still largely implicit and distributed and some functionality is also offloaded to the periphery. Robots, on the other hand, rely almost exclusively on their body models when planning and executing behaviors. I analyze the pros and cons of these different approaches and propose what may be the best solution for robots of the future. 
}

\section{Introduction}

In artificial intelligence and robotics, models of the world have been and largely still are the key means of realizing interaction of a mechanism with its environment. This position was attacked by \citet{brooks1990elephants} stating: ``The key observation is that the world is its own best model. It is always exactly up to date. It always contains every detail there is to be known. The trick is to sense it appropriately and often enough.''  \citet{brooks1991intelligence} added ``When we examine very simple level intelligence we find that explicit representations and models of the world simply get in the way.'' 

If this were the case for the world, how about for the body of an agent---human, animal, or robot? Our body seems to be even more ``always there'' than our environment.  The representationalist stance typical of robotics and (good old-fashioned) artificial intelligence \citep{haugeland1985artificial} is also applied to the body. Indeed, traditional robots heavily rely on internal models of their bodies. These are in particular the models of their kinematics---joints and links, their dimensions and orientations---and their dynamics which deal with masses and forces needed to generate motion (see 10.2.3 in \citep{Hoffmann_BodyModelsOUP_2021} for more details). With traditional robots, the interaction with the world is mediated by these models. In cognitive science, this approximately corresponds to the ``body in the brain'' approach which emphasizes representations of the body in the cerebral cortex---see for example Section 5.1 in \citep{devignemont2018mind}. This contrasts with the ``brain in the body'' or ``body in the world'' perspective, also called the sensorimotor approach that is in line with Brooks' perspective (see Section 4.2 in \citep{devignemont2018mind} or a discussion in the Introduction to \citep{ataria2021body}).

This work draws on \citep{Hoffmann_BodyModelsOUP_2021,hoffmann2022biologically,HoffmannMueller_2017}.   

\section{Biology -- from octopus to humans}
The octopus constitutes an interesting case. Belonging to cephalopods, highly derived molluscs, it is the most intelligent among them and with the largest nervous system. Cephalopods, the most advanced invertebrate class, feature, on one hand, the highest centralization of the nervous system. On the other hand, next to the central nervous system (CNS) composed of the brain and two optic lobes, there is a large peripheral nervous system (PNS) of the body and the arms. Despite the high level of centralization and in contrast to vertebrate and insect brains, there is no obvious somatotopic arrangement in either motor or sensory areas (see \citep{zullo2011new} for more details). The octopus also has a unique embodiment---a flexible body and eight arms with virtually infinite degrees of freedom. From an engineering perspective, modeling and controlling such a body (\textit{plant} in engineering jargon) using inverse kinematics and dynamics would be a nightmare. However, \citet{yekutieli2005dynamic} speculate that the octopus reaches toward a target using the following strategy: (1) Initiating a bend in the arm so that the suckers point outward. (2) Orienting the base of the arm in the direction of the target or just above it. (3) Propagating the bend along the arm at the desired speed by a wave of muscle activation that equally activates all muscles along the arm. (4) Terminating the reaching movement when the suckers touch the target by stopping the bend propagation and thus catching the target. A big part of the complexity is thus ``off-loaded'' from to the peripheral nervous system and the body itself. 

In humans, central representations of the body in the cerebral cortex certainly exist. There has been more than a century of empirical observations and theorizing, leading to concepts like body image (system of perceptions, attitudes, and beliefs pertaining to one’s own body) and body schema (system of sensory-motor capacities that function without awareness or the necessity of perceptual monitoring) (definitions taken from the Introduction to \citep{ataria2021body}). The most well-known body maps are the somatotopic representations (the ``homunculi'') in the primary motor and somatosensory cortices \citep{leyton1917observations,Penfield1937}. Yet, the somatosensory homunculi are only an ``entry point'' or ``relay station'' to downstream cortical processing rather than accurate representations or models of the body (e.g., \citep{longo2010implicit}). Downstream areas in the posterior parietal cortex (like Brodmann area 5) are thought to be involved in higher-level more integrated representations related to the configuration of the body in space, for example, but detailed understanding is still missing. Reaching in primates bears some similarity to that in the octopus. A reaching movement has some high-level characteristics like the direction of a hand's movement in space, the extent of the movement (amplitude), the overall duration (movement time), and other parameters such as anticipated level of resistance to the movement \citep{schoner2018reaching}. Also, movement generation involves cooperation between the CNS and PNS. The exact mechanisms of motor control in humans and other primates are still debated. Compared to invertebrates, motor control in vertebrates, specifically mammals and in particular primates, becomes more ``cortical'' and the motor cortex has the possibility of more direct control over the details of a particular movement, which is likely correlated with the need for dexterous manipulation (see 10.2.2 in \cite{Hoffmann_BodyModelsOUP_2021} for more details).

\section{Body models for controlling movements}
\begin{figure*}[htb]
	\centering
	\includegraphics[width=0.9\textwidth]{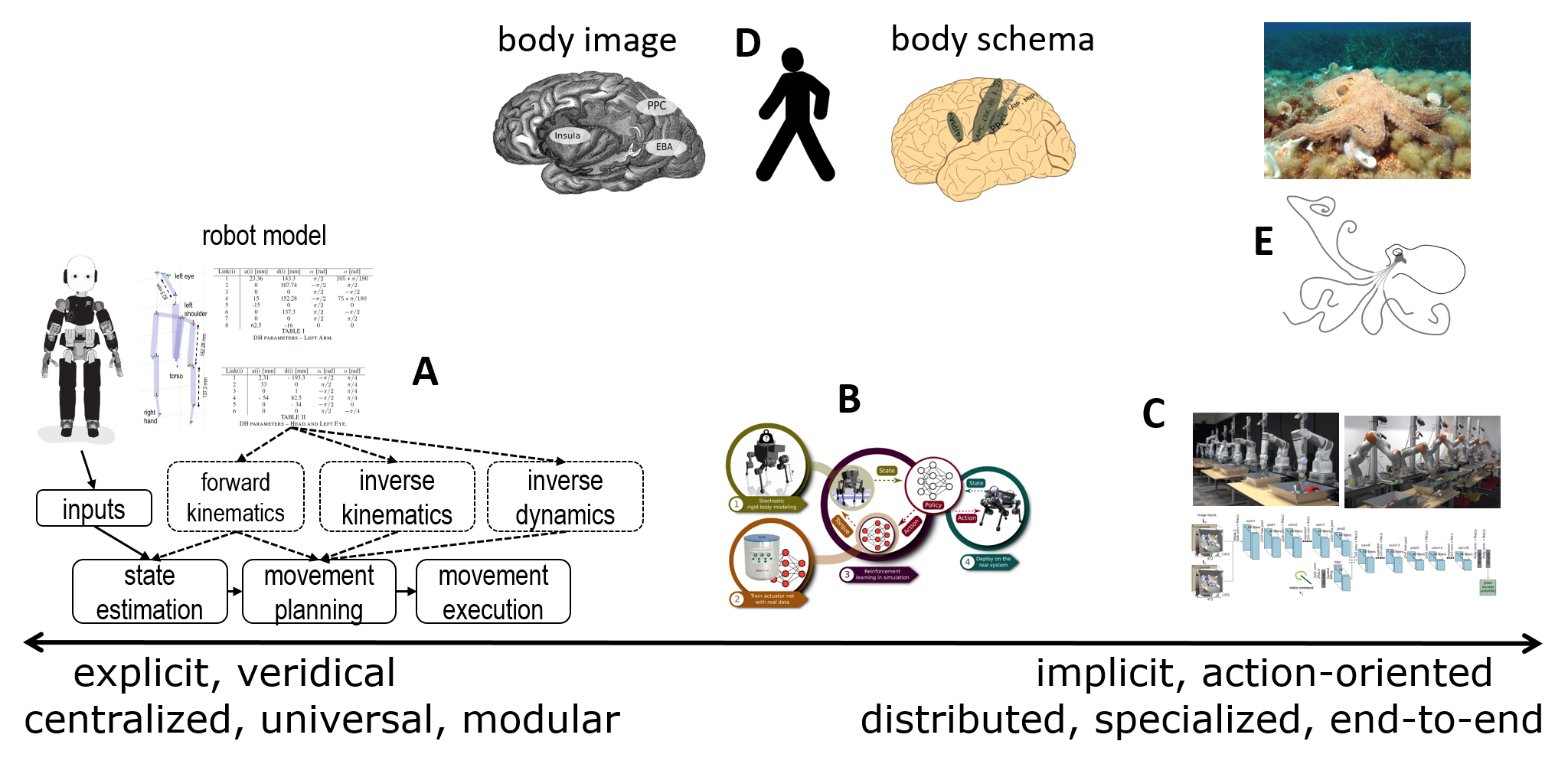}
	\caption{Body model characteristics. Upper row: examples from biology. Lower row: examples from robotics. (A) iCub humanoid robot and its models. (B) Hybrid model of the ANYmal robot \citep{hwangbo2019learning}. (C) Robot manipulators learning to grasp end-to-end \citep{levine2018learning}. (D) Human and schematic illustration of brain areas important for body representations. Brain areas involved in body image representation after \cite{berlucchi2010body}. (E) Octopus and schematic of its nervous system. \\
	Credit: A -- iCub cartoon: Laura Taverna, Italian Institute of Technology. Credit: D -- Walking human:  Public domain (https://commons.wikimedia.org/wiki/File:BSicon\_WALK.svg). Credit D -- Brain image source: Hugh Guiney / Attribution-ShareAlike 3.0 Unported (CC BY- SA 3.0).  Credit: E -- Common octopus -  albert kok / CC BY-SA (https://creativecommons.org/licenses/by-sa/3.0). Credit: E -- Octopus nervous system - Jean-Pierre Bellier / CC BY-SA (https://creativecommons.org/licenses/by-sa/4.0).}
	\label{fig:model_characteristics}
\end{figure*}

Body models can be classified according to different characteristics, such as fixed vs. adaptive, amodal vs. modal, explicit vs. implicit, serial vs. parallel, modular vs. holistic, or centralized vs. distributed \citep{Hoffmann_BodyModelsOUP_2021}. For this article, we focus on the dimensions shown in Fig.~\ref{fig:model_characteristics}.   

\subsection{Explicit and veridical versus implicit and action-oriented}
Traditional robot body models are explicit; it is clear what in the model corresponds to what in the body (e.g., a certain parameter to the length of the left forearm). They are also objective and veridical; the parameters should be the true physical values of the quantities (lengths, angles, masses, etc.). This is illustrated by the iCub humanoid robot \citep{metta2010icub} and its models positioned at the far left in Fig.~\ref{fig:model_characteristics} A. In the biological realm, representations in general are not like that and this should hold for representations of the body as well. ``What the nervous system needs to do, in general, is to transform the input into the right action'' \citep{webb2006transformation}---hence the implicit and action-oriented character of the representations. The octopus---with no known map of its body in its central nervous system---is positioned at the opposite end of the spectrum (Fig.~\ref{fig:model_characteristics} E). 
Successful action is also the only criterion for the ``quality'' of what is represented about the animal’s body in its brain; there is no need for any objective or veridical representation. Similar arguments hold for primate brains, but to a lesser extent. Numerous sites dedicated to representing the body were found (e.g., \cite{kanayama2021triadic} for a review). Compared to the octopus, much more of the body seems more explicitly represented. \citet{longo2015implicitAndExplicit} considers the implicit–explicit axis within human body representations and draws a line roughly between the body schema and the body image. In tasks more related to action and where humans do not consciously represent their body, the body models seem more implicit and also less accurate. These representations may also be dominated by somatosensation and inherit some of the distortions typical of the somatosensory homunculi. Conversely, tasks that relate to conscious perception of our body seem to draw on more explicit representations that are also more accurate/veridical (e.g., image of our hand). This is schematically illustrated in Fig.~\ref{fig:model_characteristics}, D.

\subsection{Centralized, universal, modular versus distributed, specialized, end-to-end}

Robot models are normally centralized---exist only in one place in the robot software. On the other hand, neural representations are known to be distributed. Whereas this ``spatial aspect'' may be also related to the computational substrate (computers versus neurons), more important is a functional division. Albeit centralized, robot body models are highly \textit{modular}. For the iCub (Fig.~\ref{fig:model_characteristics} A), there would normally be a single model of its kinematics and another one of its dynamics (mass distribution etc.). Then, there are distinct modules like forward/inverse kinematics and dynamics that may draw from the same robot model and be recruited for different purposes like state estimation, movement planning etc. There would be typically only one module of every kind (imagine a software library) providing this functionality upon request. The representations/modules will thus be \textit{universal} and not overlapping. For deep learning applied to robotics, this is not the case. \citet{levine2018learning} specialize on a single task (grasping); a different task will likely need a different network. The representations are thus end-to-end, task-specific and in case of multiple tasks also overlapping. In nervous systems, there may also be complete sensorimotor loops specialized on individual tasks, partially overlapping or redundant. However, this approach does not scale well. In primates, the posterior parietal cortex is regarded as a site where information about the body from different modalities converges. Specific areas related to representations of body parts or reaching targets in different reference frames have been found. These are recruited in different tasks or contexts and hence, there is certain universality and modularity---again more for the body image than body schema Fig.~\ref{fig:model_characteristics}, D.

\section{Use the body directly}
\label{sec:use_body_directly}

While body representations can take very different forms, one should at the same time consider the radical possibility of using the body directly rather than through an internal model. Again, as \citet{brooks1990elephants} put it: ``The key observation is that the world is its own best model. It is always exactly up to date. It always contains every detail there is to be known. The trick is to sense it appropriately and often enough.'' In fact, for the case of the body, one can do even without sensing its state. First, there are examples how a completely passive body, ``pure physics'', can generate useful behavior. In the biological realm, this is for example the body of a trout. \citet{liao2004neuromuscular} shows that, paradoxically, under specific circumstances, a dead trout body can exploit vortices in the water to the extent that it swims upstream. In robotics, a similar well-known example are the passive dynamics walkers \cite{McGeer1990}---carefully designed mechanical devices that walk down a ramp without any motors, sensors, or controllers. Second, in case there is actuation, we have privileged access to the body current or future state (using forward models / efference copies \citep{Webb2004}) and hence, it may be unnecessary to sense it. 

\begin{figure*}[htb]
	\centering
	\includegraphics[width=0.9\textwidth]{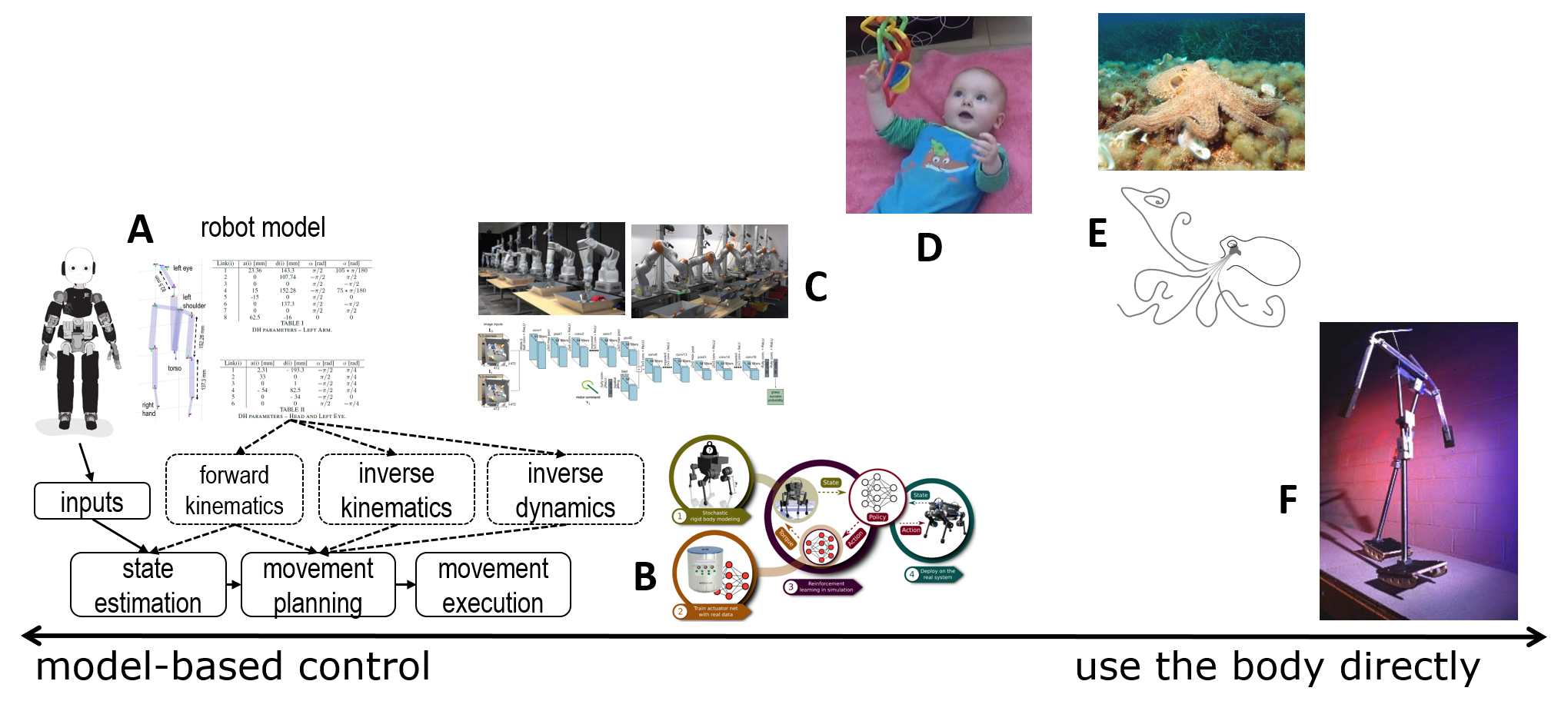}
	\caption{Model-based control or direct use of the body.  (A) iCub humanoid robot and its models. (B) Model of the ANYmal robot \citep{hwangbo2019learning}. (C) Robot manipulators learning to grasp end-to-end \citep{levine2018learning}. (D) Infant reaching. (E) Octopus and schematic of its nervous system. (F) The Cornell passive biped with arms \cite{Collins2005}.\\
	Credit: A, E  -- see Fig.~\ref{fig:model_characteristics}. Credit F: H. Morgan.}
	\label{fig:model_or_no_model}
\end{figure*}

Different positions on the imaginary landscape ranging from model-based control to direct use of the body are illustrated in Fig.~\ref{fig:model_or_no_model}. The passive dynamic walker is positioned at the far right of the schematics. As discussed above, the octopus is able to reach for visual targets, but it may not know---and may not need to know---how long its arm is or where it is exactly in space. Orienting the base of the arm and propagating the bend until contact is detected by the suckers may well suffice. The need to represent the body, its state, and the complex inverse kinematics and dynamics has been largely offloaded to embodiment---the properties of the octopus arm, supported by the peripheral nervous system and low-dimensional inputs from the central nervous system. Human reaching, Fig.~\ref{fig:model_or_no_model} D, is probably less embodied compared to the octopus, but still sharing some important characteristics. \citet{cisek2003reaching} highlight the importance of online, dynamically generated character of movement generation in primates. At the same time, they also point out that due to conduction delays inherent to the sensorimotor system, purely feedback control is limited, or at least slow. Thus, feed-forward commands and local neural reflex loops have to work in concert. Robots, on the other hand, typically heavily rely on models. Importantly, this is the case also for the solutions employing deep learning. In \citep{levine2018learning}, Fig.~\ref{fig:model_or_no_model} C, the embodiment of the robot arm or the gripper is not significantly exploited.

\section{Robots: with or without a model?}
Mechanical engineers naturally think in terms of how to make the best design of a machine for a task. However, control engineers have a strong preference for model-based control. Moreover, solutions for nonlinear systems are much more difficult to obtain, and they often involve a linearization of the system of some sort. Thus, complex (highly dimensional, dynamic, nonlinear,
compliant, deformable, ‘soft’) robot bodies are avoided as they cannot be modeled and controlled with the available methods. Many robot engineers then simply take the body as fixed and seek to exploit to the maximum what can be done at the ``software level''.

Including the parameters of the body into the design considerations may give rise to better performance of the whole system; these may be solutions involving a simpler controller, but also solutions that were previously unattainable when the body was fixed. Following the dynamical systems' perspective, \citet{Fuchslin2013} provide an illustration of the possible goals of the design process: (1) To design the physical dynamical system such that desired regions of the state space have attracting properties. Then it is sufficient to use a simple control signal that will bring the system to the basins of attraction of individual stable points that correspond to target behaviors. (2) More complicated behavior can be achieved if the attractor landscape can be manipulated by the control signal. 

If a mathematical formulation of the controller and the plant is available, this design methodology can be directly applied. The first part is demonstrated by on the passive dynamic walker \citep{McGeer1990}: The influence of scale, foot radius, leg inertia, height of center of mass, hip mass and damping, mass offset, and leg mismatch is evaluated. In addition, the stability of the walker is calculated. Jerrold Marsden and his coworkers presented a method that allows for co-optimization of the controller and plant by combining an inner loop (with discrete mechanics and optimal control) and an outer loop (multiscale trend optimization). They applied it to a model of a walker and obtained the best position of the knee joints (\citep{Pekarek2010} -- Ch. 5).
However, typical real-world agents are more complex than simple walkers. \citet{HolmesFull2006} provide an excellent dynamical systems analysis of the locomotion of rapidly running insects and derive implications for the design of the RHex robot. Yet, they conclude that ``a gulf remains between the performance we can elicit empirically and what mathematical analyses or numerical simulations can explain. Modeling is still too crude to offer detailed design insights for dynamically stable autonomous machines in physically interesting settings.'' Modeling and optimization of more complicated morphologies---like compliant structures---is nevertheless an active research topic (e.g., \citep{Wang2009}). The second point of \citet{Fuchslin2013}---achieving ``morphological programmability'' by constructing a dynamical system with a parametrized attractor landscape---remains even more challenging though. 

One of the merits of exploiting the contributions of body morphology should be that the physical processes do not need to be modeled, but can be used directly. However, without a model of the body at hand, several body designs need to be produced and---together with the controller---tested in the respective task setting. The design space of the joint controller-body system blows up and we may be facing a curse of dimensionality. This is presumably the strategy adopted by the evolution of biological organisms that could cope with the enormous dimensionality of the space. In robotics, this has been taken up by evolutionary robotics \citep{Nolfi2000}. The simulated agents of  \citet{Sims1994} demonstrate that co-evolving brains and bodies together can give rise to unexpected solution to problems. \citet{Bongard2011} showed that morphological change indeed accelerates the evolution of robust behavior in such a brain-body co-evolution setting. With the advent of rapid prototyping technologies, physics-based simulation could be complemented by testing in real hardware \citet{Lipson2000}, but this reintroduces the modeling through the back door: the phenotypes in the simulator now become models and they need to sufficiently match their real counterparts. Yet, a ``reality gap'' \citep{Jakobi1995,Koos2013} always remains between simulated and real physics. The only alternative is to optimize in hardware directly, which is in general slow and costly.  \citet{Brodbeck2015} provide an interesting illustration how locomoting cube-like creatures can be evolved in a model-free fashion through automated manufacturing and testing. However, in summary, the design decisions---which parameters to optimize---are based on heuristics and a clear methodology is still missing. Furthermore, with the absence of an analytical model of the controller and plant, no guarantees on the system's performance can be given.

\section{Conclusion and outlook}
Rich properties of complex bodies (highly dimensional, dynamic, nonlinear, compliant and deformable) have been mostly overlooked or deliberately suppressed by classical mechatronic designs, as they are largely incompatible with traditional control frameworks, where linear plants are preferred. This is definitely a missed opportunity. On the other hand, while complex bodies carry a lot of ``self-control'' potential, this property does not come for free. It has to be said that the exploitation of truly complex bodies to accomplish tasks is still mostly at a ``proof-of-concept'' stage. A closely connected issue is the one of modeling of these systems---complex, or for example soft, bodies are notoriously difficult to model. The model may not be necessary for the system to perform the task; however, without a model, the understanding and design is more complicated and performance guarantees are limited. The field, which has been dominated by heuristics so far, needs to embrace more systematic approaches that allow to navigate in this complex landscape.

The area of soft robotics and morphological computation/morphological control/morphology facilitating control \citep{Fuchslin2013,Mueller2017}  is rife with different trading spaces \citep{Pfeifer2013}. As we move from the traditional engineering framework with a central controller that commands a ``dumb'' body toward delegating more functionality to the physical morphology, some convenient properties will be lost. In particular, the solutions may not be portable to other platforms anymore, as they will become dependent on the particular morphology and environment (the passive dynamic walker is the extreme case). The versatility of the solutions is likely to drop as well. To some extent, the morphology itself can be used to alleviate these issues---if it becomes adaptive. Online changes of morphology (like changes of stiffness or shape) thus constitute another tough technological challenge. Completely new, distributed control algorithms that rely on self-organizing properties of complex bodies and local distributed control units will need to be developed \citep{McEvoy2015,Rieffel2010}. 

In summary, computer scientists, roboticists, and control engineers impose a representationalist perspective on designing machines and their behaviors. This is similar to traditional cognitive science (cognitivism). It is sometimes acknowledged that the representations---world or body---should be embodied. However, rather than ``embodied body models'', it seems more natural to think of the ``brain in the body'' or ``body in the world'' (cf. discussion in \citep{alsmith2012embodying,devignemont2018mind,ataria2021body}, and more direct use of the body wherever possible. For engineers, this will be a major challenge though.


\vspace{5ex}
\noindent
{\bf\large Acknowledgement}\vspace{2ex} \newline
{ This work was supported by the Czech Science Foundation (GA CR), project no. 20-24186X. I would like to thank to Rolf Pfeifer for discussions along these lines.
}

\vspace{3,16314mm}
\nocite{*}
\bibliographystyle{apa}
\bibliography{Hoffmann_BodySchema}


\end{document}